\documentclass[%
 reprint,
superscriptaddress,
 amsmath,amssymb,
 aps,
 pre, 
]{revtex4-2}

\usepackage{graphicx}
\usepackage{dcolumn}
\usepackage{bm}
\usepackage{orcidlink}

\begin{document}

\preprint{APS/123-QED}

\title{Statistical mechanics of a three-body Hopfield model with finite connectivity}

\author{Yushi Sugawara}
\affiliation{
Department of Integrated Sciences,
The University of Tokyo,
 Komaba, Meguro-ku, Tokyo 153-8902, Japan
}
\author{Koji Hukushima\orcidlink{0000-0003-1153-1758}}%
\affiliation{
 Graduate School of Arts and Sciences,
 The University of Tokyo,
 Komaba, Meguro-ku, Tokyo 153-8902, Japan
}
\affiliation{
Komaba Institute for Science, The
University of Tokyo, 3-8-1 Komaba, Meguro-ku, Tokyo 153-8902, Japan
}

\date{\today}

\begin{abstract}
A three-body Hopfield model defined on sparse random graphs with finite mean connectivity is studied using a replica-symmetric (RS) analysis. By extending the functional replica framework for conventional two-body finite-connectivity Hopfield models, self-consistent equations for the local-field distribution are derived and solved numerically by population dynamics. 
At low memory load, the RS analysis shows a discontinuous transition between paramagnetic and retrieval states, accompanied by coexistence of distinct fixed points and an upper spinodal of the retrieval solution.  
Starting the RS population dynamics from an uninformative finite-amplitude field distribution also leads to a spin-glass-like (SG-like) fixed point at low temperatures, even though the paramagnetic fixed point remains locally stable. As the number of embedded patterns increases, the retrieval solution disappears, while the SG-like RS solutions are found at high load. However, the local instability and thermodynamic inconsistencies of the low-temperature RS solutions indicate the limitations of the RS description of the glassy regime. 
Finite-size Monte Carlo simulations using population annealing agree well with the RS description of the retrieval branch during heating but exhibit glassy trapping behavior under cooling. Such trapping can occur even for a single embedded pattern, where conventional cross-talk noise among stored patterns is absent, indicating the importance of sparse connectivity and three-body interaction.  
The characteristic number of embedded patterns separating different solution regimes grows approximately linearly with the mean connectivity, corresponding to finite characteristic load ratios. These results clarify how sparse connectivity and many-body interactions influence the stability and accessibility of retrieval states in associative memory.  
\end{abstract}

\maketitle


\section{Introduction}
\label{sec:introduction}
The Hopfield model is a fundamental paradigm of associative memory, in which neuronal states are represented by Ising spins and memory patterns are embedded through Hebbian interactions~\cite{hopfield1982Neural, Amari1972, Little1974101}. For the standard fully connected model with two-body interactions, equilibrium statistical mechanics has successfully clarified the phase diagram in the temperature-load plane. This phase diagram consists of retrieval, spin-glass, and paramagnetic phases, and yields the critical storage capacity of the model~\cite{AGS1985a, AGS1985b}.

One notable extension to the standard model is coupling dilution, which is partly motivated by the finite connectivity of biological neural networks. Let $N$ be the total number of neurons and $c$ the mean connectivity per neuron. In the dense regime where $c=\mathcal{O}(N)$, the storage capacity and the information content per coupling of partially connected networks have been extensively analyzed~\cite{canning1988Partially}. In the extremely diluted limit, where $c\to\infty$ while $c/N\to 0$, the dynamics of asymmetric networks~\cite{derrida1987Exactly} and the equilibrium phase diagram of symmetric networks~\cite{watkin1991Neural} have been established. In contrast, in the finite-connectivity regime, where $c=\mathcal{O}(1)$, the local field is composed of a finite number of random contributions. As a result, the macroscopic state must be described by replica order-parameter functions rather than by a finite number of scalar order parameters. The development of replica and cavity methods for finite-connectivity spin glasses and sparse random systems has provided a powerful framework for analyzing such models~\cite{VianaBray1985, Kanter_Sompolinsky_1987, Mezard_Parisi_2001, MezardMontanari}. For the symmetric two-body Hopfield model with finite connectivity, self-consistent equations for the local-field distribution have been explicitly derived, and it has been shown that, within the replica-symmetric (RS) ansatz, all phase transitions separating the paramagnetic, retrieval, and spin-glass phases are continuous~\cite{wemmenhove2003Finite}. 

Another important direction is the study of associative memory models with many-body interactions. Such models have been investigated since the mid-1980s in relation to storage capacity, the number of metastable states, thermodynamics, and basins of attraction~\cite{Peretto1986, Baldi_etal_1987, gardner1987Multiconnected, AbbotArian1987, HornUsher_1988, E_Gardner_1989}. In a representative fully connected $r$-body Hebbian model, the number of storable patterns scales as $\mathcal{O}(N^{r-1})$~\cite{AbbotArian1987}, which is a substantial enhancement compared with the $\mathcal{O}(N)$ scaling of the two-body model. This enhancement, however, comes at the cost of increasing the number of independent coupling parameters to $\mathcal{O}(N^{r})$, leading to a polynomial increase in the computational cost of retrieval updates. Moreover, higher-order interactions are known to generate a proliferation of metastable states, which can shrink the basins of attraction for memory states~\cite{HornUsher_1988, E_Gardner_1989}. Thus, an increase in storage capacity does not necessarily imply an improvement in practical retrieval performance. A further important feature of models with $r>2$, closely related to many-body spin glasses, is the appearance of discontinuous phase transitions~\cite{gardner1987Multiconnected, Baldi_etal_1987}. In such cases, non-trivial ordered phases can emerge while the paramagnetic solution remains locally stable, giving rise to pronounced hysteresis and strong initial-condition dependence. Glassy behavior in sparse many-body systems has also been investigated in finite-connectivity spin glasses and related optimization problems, including studies of one-step replica-symmetry-breaking solutions for diluted $p$-spin glasses and random $K$-XORSAT~\cite{Franz-PRL2001}. Glassy trapping under cooling from random initial conditions has also been reported in finite-connectivity three-spin ferromagnets~\cite{S.Franz2001}. 

Recently, renewed interest in many-body associative memory has been stimulated by progress in modern Hopfield networks and their connection to the attention mechanism in Transformers~\cite{ramsauer2021hopfield,krotov2016Dense}. From an engineering standpoint, modern Hopfield models with continuous degrees of freedom and gradient-based retrieval dynamics provide a powerful framework for optimizing storage capacity and computational efficiency. Generalized Hopfield models with nonlinear interactions have been shown to achieve storage capacity that grows exponentially with the number of neurons~\cite{Demircigil2017}. More recently, statistical-mechanical analyses of dense associative memories have characterized their retrieval thresholds and complex phase diagrams in the fully connected regime~\cite{LucibelloMezard_2024}. These developments often combine several ingredients: many-body interactions, continuous state space, and gradient-based retrieval dynamics. From a fundamental standpoint, however, this combination makes it difficult to distinguish the role played by the algebraic structure of many-body interactions from the effects of continuous variables, which can smooth the energy landscape and modify trapping behavior. To clarify the intrinsic physical consequences of higher-order interactions under constrained architectures, it is useful to examine a benchmark limit, namely, a discrete spin model defined on a sparse graph with a finite mean connectivity $c=\mathcal{O}(1)$. In this sparse regime, the absolute number of storable random patterns is expected to remain finite in the thermodynamic limit because the number of nonzero couplings is only $\mathcal{O}(N)$. Nevertheless, this limit provides a clean setting for studying how discontinuities and metastability intrinsic to many-body interactions compete with the structural disorder of a sparse graph, without the smoothing effects of continuous degrees of freedom. 

In this study, we extend the statistical-mechanical framework of finite-connectivity two-body Hopfield models~\cite{wemmenhove2003Finite} to symmetric Hebbian three-body interactions distributed over a sparse random triplet structure. Our aim is to clarify how higher-order interactions modify the phase structure and retrieval properties of associative-memory models with a finite number of stored patterns. Motivated by the limited number of nonzero couplings in this regime, we focus on a finite number $P=\mathcal{O}(1)$ of embedded memory patterns in the thermodynamic limit $N\to\infty$. We define the normalized load parameter $\alpha=P/c$. Since $P$ is an integer, $\alpha$ takes discrete values for fixed $c$. 
For $P=1$, the model is gauge-equivalent to a finite-connectivity three-spin ferromagnet, providing a direct connection to the study by Franz et al.~\cite{S.Franz2001}.

Following Wemmenhove and Coolen~\cite{wemmenhove2003Finite}, we employ the replica method combined with the pattern-sublattice technique to derive the RS free-energy functional and self-consistency equations for the local-field distribution. The same equations and corresponding RS thermodynamic quantities can also be obtained by using the cavity method, which provides a more direct formulation in terms of local messages on locally tree-like random hypergraphs~\cite{Mezard_Parisi_2001}. In this interpretation, the local field can be interpreted as a sum of three-body messages whose number follows a Poisson distribution. This structure also provides the basis for solving the self-consistent equations numerically by the population dynamics (PD) method. 

Our analysis reveals that, in the three-body finite-connectivity model, the paramagnetic and ordered retrieval solutions coexist in the low-load regime, with a clear spinodal structure and initial-condition-dependent hysteresis. In the PD analysis, the retrieval solution remains stable up to its spinodal temperature when the iteration is initialized close to an embedded pattern. In contrast, when uninformative random initial conditions are used, the iteration converges to a spin-glass-like fixed point at low temperatures. 
We further compare the RS PD results with finite-size Markov-chain Monte Carlo (MCMC) simulations using population annealing (PA)~\cite{HukushimaIba, Machta2010}. The heating protocol quantitatively reproduces the RS retrieval branch, with finite-size deviations near its upper spinodal, whereas the cooling protocol exhibits pronounced hysteresis and glassy trapping. Such trapping was previously observed in finite-connectivity three-spin ferromagnets~\cite{S.Franz2001}, which are closely related to the single-pattern limit of our model. Our results demonstrate that similar trapping behavior persists when multiple memory patterns are embedded, suggesting that it cannot be attributed only to the conventional cross-talk mechanism. Furthermore, by varying the memory load, we characterize how the retrieval and SG-like fixed point structures change within the RS description. These findings clarify how finite connectivity and many-body interactions determine the phase structure and accessibility of associative memories. 

The rest of this paper is organized as follows. In Sec.~\ref{sec:model}, we introduce the three-body Hopfield model on a sparsely connected graph and define the normalized load parameter $\alpha$. In Sec.~\ref{sec:free_energy}, we evaluate the typical free-energy density using the replica method. By employing the pattern-sublattice representation and the RS ansatz, we derive the self-consistent functional equations for the local-field distribution and formulate the RS stability conditions. In Sec.~\ref{sec:results}, we solve the self-consistent equations using the PD method and compare the results with population annealing Monte Carlo simulations. We then discuss the temperature dependence of the order parameters and thermodynamic quantities, present the RS phase diagram in the $\alpha$-$T$ plane, and examine the mean-connectivity dependence of the characteristic load parameters. Finally, Sec.~\ref{sec:summary} is devoted to the summary and discussion of our findings.  

\section{Model}
\label{sec:model}
We consider a system of $N$ Ising spins, $\bm{\sigma}=(\sigma_1,\ldots,\sigma_N)\in\{-1,1\}^N$. Let $\bm{\Xi}=\{\bm{\xi}^\mu\}_{\mu=1}^P$ denote $P$ random patterns embedded in the system, where each pattern is given by $\bm{\xi}^\mu=(\xi_1^\mu,\ldots,\xi_N^\mu)\in\{-1,1\}^N$. The Hamiltonian with three-body interactions is defined as 
\begin{equation}
    H(\bm{\sigma}) = -\sum_{i<j<k}J_{ijk}\sigma_i\sigma_j\sigma_k, 
\end{equation}
where $J_{ijk}$ denotes the interaction strength among spins $i$, $j$, and $k$. The couplings are given by a three-body generalization of the Hebbian learning rule, 
\begin{equation}
    J_{ijk} = \frac{c_{ijk}}{c}\sum_{\mu=1}^P\xi_i^\mu\xi_j^\mu\xi_k^\mu, 
    \label{eqn:jij}
\end{equation}
where $c_{ijk}\in\{0,1\}$ are independent quenched random variables describing the connectivity, and $c$ represents the mean connectivity per spin. 

The pattern components $\xi_i^\mu$ are independent quenched random variables satisfying  
\begin{equation}
    \mathbb{P}(\xi_i^\mu = \pm 1) = \frac{1}{2}.
    \label{eqn:pattern_distribution}
\end{equation}
We denote the expectation with respect to the pattern distribution by $\langle\cdots\rangle_{\bm{\xi}}$. In particular, 
\begin{equation}
    \langle \xi_i^\mu\rangle_{\bm{\xi}}=0, \quad \langle \xi_i^\mu\xi_i^\nu\rangle_{\bm{\xi}}=\delta_{\mu\nu}. 
\end{equation}
The distribution of the connectivity variables $\bm{c}=\{c_{ijk}\}$ is given by 
\begin{equation}
    \mathbb{P}(\bm{c}) = \prod_{i<j<k}\left(\frac{2c}{N^2}\delta_{c_{ijk},1} + \left(1-\frac{2c}{N^2}\right)\delta_{c_{ijk},0}\right),
    \label{eqn:c_distribution}
\end{equation}
where $c_{ijk}=c_{jki}=c_{kij}$ and $c_{ijk}=0$ whenever two indices coincide. With this choice, the expected number of connected triplets containing a given site $i$ is 
\begin{equation}
    \sum_{j<k}\overline{c_{ijk}} = \binom{N-1}{2}\frac{2c}{N^2}\simeq c.
\end{equation}
Here, $\overline{\cdots}$ denotes the average over the connectivity variables $\bm{c}$ according to Eq.~(\ref{eqn:c_distribution}). Equivalently, the total number of connected triplets is $Nc/3$ on average, since each connected triplet contains three spins. 

In this study, we focus on the finite-connectivity regime, where $c=\mathcal{O}(N^0)$. Since the stored information is encoded in the couplings rather than in the spin variables themselves, the number of storable random patterns is limited by the number of independent interaction parameters. In the present sparse regime, the total number of nonzero couplings scales as $\mathcal{O}(N)$. By analogy with the known storage-capacity scaling in the fully connected higher-order Hopfield models~\cite{AbbotArian1987}, this suggests that the number of storable patterns remains finite in the thermodynamic limit $N\to\infty$. We, therefore, introduce the normalized load parameter 
\begin{equation}
    \alpha = \frac{P}{c}. 
\end{equation}

\section{Free energy and self-consistent equations}
\label{sec:free_energy}
In this section, we study the thermodynamic properties of the model using the replica method~\cite{MPV1987, MezardMontanari}. We first formulate the quenched free-energy density in the thermodynamic limit for a fixed typical realization of the embedded patterns. We then introduce sublattice-resolved order-parameter functions and derive the self-consistent equations. Finally, by imposing the replica symmetry (RS) ansatz, we obtain the saddle-point condition for the local-field distribution and the corresponding free-energy density.  

\subsection{Replica method and sublattice decomposition}
To evaluate the typical free-energy density, we employ the replica method. The embedded patterns $\bm{\Xi}$ are quenched random variables. In the following, we first average over the connectivity variables for a fixed typical realization of $\bm{\Xi}$, and then treat the dependence on the patterns through the sublattice representation introduced below.  For a given inverse temperature $\beta$ and fixed patterns $\bm{\Xi}$, the free-energy density $f$ is written as  
\begin{equation}
    f(\bm{\Xi)} = -\lim_{N\to\infty}\lim_{n\to 0} \frac{1}{\beta Nn}\ln \overline{Z^n}(\bm{\Xi},\bm{c}). 
\end{equation}
The replicated partition function before the connectivity average is  
\begin{equation}
    {Z^n}(\bm{\Xi},\bm{c}) = \sum_{\bm{\sigma}^1,\ldots,\bm{\sigma}^n}\exp\left(\beta\sum_{i<j<k}\sum_{\mu=1}^P\sum_{\alpha=1}^n J_{ijk}\sigma_i^\alpha\sigma_j^\alpha\sigma_k^\alpha\right).
\end{equation}
Substituting the definition of the coupling $J_{ijk}$ from Eq.~(\ref{eqn:jij}), the average over the independent connectivity variables gives  
\begin{equation}
    \overline{Z^n}(\bm{\Xi}) = \sum_{\bm{\sigma}^1,\ldots,\bm{\sigma}^n} \prod_{i<j<k} \overline{ \exp\left( \frac{\beta}{c} c_{ijk} \sum_{\mu=1}^P \xi_i^\mu\xi_j^\mu\xi_k^\mu \sum_{\alpha=1}^n \sigma_i^\alpha\sigma_j^\alpha\sigma_k^\alpha \right) }.
\end{equation}

 We introduce the pattern vector $\bm{\xi}_i=(\xi_i^1,\dots,\xi_i^P)$ and the replica vector $\bm{\sigma}_i=(\sigma_i^1,\ldots,\sigma_i^n)$. For notational compactness, we use the shorthand notation 
 \begin{equation}
    (\bm{\xi}_i\bm{\xi}_j\bm{\xi}_k)=\sum_{\mu=1}^P\xi_i^\mu\xi_j^\mu\xi_k^\mu, \quad 
    (\bm{\sigma}_i\bm{\sigma}_j\bm{\sigma}_k) = \sum_{\alpha=1}^n\sigma_i^\alpha\sigma_j^\alpha\sigma_k^\alpha. 
\end{equation}
Here and in what follows, the same notation is used for any three pattern vectors; for example, 
\[
(\bm{\xi}\bm{\xi}'\bm{\xi}'') = \sum_{\mu=1}^P \xi^\mu\xi'^\mu\xi''^\mu.
\]
 After averaging over the connectivity variables $\bm{c}$ and retaining the leading contribution in the large-$N$ limit, we obtain 
\begin{equation}
    \overline{Z^n}\simeq 
    \sum_{\bm{\sigma}^1,\ldots,\bm{\sigma}^n}\exp\left(\frac{c}{3N^2}\sum_{ijk}\left[e^{\frac{\beta}{c}(\bm{\xi}_i\bm{\xi}_j\bm{\xi}_k)(\bm{\sigma}_i\bm{\sigma}_j\bm{\sigma}_k)}-1\right]
    \right),
    \label{eqn:replica_pfunction}
\end{equation}
The exponent scales as $\mathcal{O}(N)$, while subleading contributions from coincident indices are at most $\mathcal{O}(N^0)$ and therefore vanish in the free-energy density. 

To handle the site-dependent pattern variables, it is convenient to introduce sublattices $I_{\bm{\xi}}$, which partition the $N$ sites according to their specific pattern vector~\cite{vanHemmen_1986}. Specifically, 
\begin{equation}
    I_{\bm{\xi}} = \{\, i \mid \bm{\xi}_i=\bm{\xi}\, \}, 
\end{equation}
is the set of sites whose $P$-component pattern vector is equal to $\bm{\xi}$. We define the relative size of each sublattice as $p_{\bm{\xi}}=|I_{\bm{\xi}}|/N$. Since the pattern vectors are drawn independently at random, the empirical distribution $p_{\bm {\xi}}$ converges to the underlying pattern distribution in the thermodynamic limit for fixed finite $P$. Consequently, for any function $g(\bm{\xi})$, the site average can be rewritten exactly as a sublattice average, which becomes self-averaging in the thermodynamic limit:   
\begin{equation}
    \frac{1}{N}\sum_{i=1}^Ng(\bm{\xi}_i) = \sum_{\bm{\xi}}p_{\bm{\xi}}g(\bm{\xi})\xrightarrow[N\to\infty]{}
    \langle g(\bm{\xi}) \rangle_{\bm{\xi}}.
    \label{eqn:self_averaging}
\end{equation}

Using this sublattice representation, the site-dependent part of the exponent in Eq.~(\ref{eqn:replica_pfunction}) can be rewritten by grouping sites with identical pattern vectors: 
\begin{align}
    \sum_{ijk}e^{\frac{\beta}{c}(\bm{\xi}_i\bm{\xi}_j\bm{\xi}_k)(\bm{\sigma}_i\bm{\sigma}_j\bm{\sigma}_k)} \nonumber 
   &=
  \sum_{\bm{\xi},\bm{\xi}^\prime,\bm{\xi}^{\prime\prime}} \sum_{\bm{\sigma},\bm{\sigma}^\prime,\bm{\sigma}^{\prime\prime}}e^{\frac{\beta}{c}(\bm{\xi}\bm{\xi}^\prime\bm{\xi}^{\prime\prime})(\bm{\sigma}\bm{\sigma}^{\prime}\bm{\sigma}^{\prime\prime})}\nonumber \\
  &\times
  \sum_{i\in I_{\bm{\xi}}} \delta_{\bm{\sigma},\bm{\sigma}_i}
  \sum_{j\in I_{\bm{\xi}^\prime}} \delta_{\bm{\sigma}^\prime,\bm{\sigma}_j}
  \sum_{k\in I_{\bm{\xi}^{\prime\prime}}} \delta_{\bm{\sigma}^{\prime\prime},\bm{\sigma}_k},\nonumber
\end{align}
where
\[
\delta_{\bm{\sigma},\bm{\sigma}_i}= \prod_{\alpha=1}^n\delta_{\sigma^\alpha,\sigma_i^\alpha}
\]
is the Kronecker delta for replicated spin vectors. 

\subsection{Order-parameter function and saddle-point equations}
Following the functional order-parameter approach for finite-connectivity disordered systems~\cite{Monasson_1998} and the pattern-sublattice representation for Hopfield-type models~\cite{vanHemmen_1986, wemmenhove2003Finite}, the macroscopic state of the system is described by the order-parameter function  $P_{\bm{\xi}}(\bm{\sigma})$. This function represents the probability distribution of the replica spin vector $\bm{\sigma}$ within the sublattice $I_{\bm{\xi}}$. We introduce this function through the identity: 
\begin{equation}
    1 = \int\prod_{\bm{\xi},\bm{\sigma}}dP_{\bm{\xi}}(\bm{\sigma}) \prod_{\bm{\xi},\bm{\sigma}}\delta\left(P_{\bm{\xi}}(\bm{\sigma})-\frac{1}{|I_{\bm{\xi}}|}\sum_{i\in I_{\bm{\xi}}} \delta_{\bm{\sigma},\bm{\sigma}_i}\right).
    \label{eqn:identity}
\end{equation}
Here, $\delta(\cdots)$ denotes the ordinary Dirac delta function applied to each component $P_{\bm{\xi}}(\bm{\sigma})$. 
By substituting Eq.~(\ref{eqn:identity}) into the expression of the replicated partition function and using the integral representation of the $\delta$-function with the conjugate function $\hat{P}_{\bm{\xi}}(\bm{\sigma})$, the replicated partition function $\overline{Z^n}$ averaged over the connectivity variables is expressed as  
\begin{equation}
    \overline{Z^n} = \int\prod_{\bm{\xi},\bm{\sigma}}dP_{\bm{\xi}}(\bm{\sigma})d\hat{P}_{\bm{\xi}}(\bm{\sigma})\exp\left(N\phi\left[\{P_{\bm{\xi}},\hat{P_{\bm{\xi}}}\}\right]\right),
\end{equation}
where the functional action $\phi$ is given by  
\begin{widetext}
\begin{equation}
    \phi  =  i \sum_{\bm{\sigma}} \left\langle\hat{P}_{\bm{\xi}}(\bm{\sigma}){P}_{\bm{\xi}}(\bm{\sigma})\right\rangle_{\bm{\xi}} + \left\langle\ln\sum_{\bm{\sigma}}e^{-i\hat{P}_{\bm{\xi}}(\bm{\sigma})}\right\rangle_{\bm{\xi}} +\frac{c}{3}\left\langle
    \sum_{\bm{\sigma},\bm{\sigma}^\prime,\bm{\sigma}^{\prime\prime}}P_{\bm{\xi}}(\bm{\sigma})P_{\bm{\xi}^\prime}(\bm{\sigma}^\prime)P_{\bm{\xi}^{\prime\prime}}(\bm{\sigma}^{\prime\prime})\left(e^{\frac{\beta}{c}(\bm{\xi}\bm{\xi}^\prime\bm{\xi}^{\prime\prime})(\bm{\sigma}\bm{\sigma}^\prime\bm{\sigma}^{\prime\prime})}-1\right)
    \right\rangle_{\bm{\xi},\bm{\xi}^\prime,\bm{\xi}^{\prime\prime}}.
    \label{eqn:action}
\end{equation}
\end{widetext}
In the thermodynamic limit $N\to\infty$, the functional integral is evaluated by the saddle-point method. The quenched free-energy density is given by
\begin{equation}
    f(\bm{\Xi}) = -\lim_{n\to 0}\frac{1}{\beta n}\mathrm{Extr}_{\{P_{\bm{\xi}},\hat{P}_{\bm{\xi}}\}}\phi\left[\{P_{\bm{\xi}},\hat{P_{\bm{\xi}}}\}\right], 
    \label{eqn:free_energy_expression}
\end{equation}
where $\mathrm{Extr}$ denotes extremization with respect to the functions $P_{\bm{\xi}}(\bm{\sigma})$ and $\hat{P}_{\bm{\xi}}(\bm{\sigma})$. 

Taking the functional variation of $\phi$ with respect to ${P}_{\bm{\xi}}(\bm{\sigma})$ yields the saddle-point condition, 
\begin{widetext}
\begin{equation}
    i\hat{P}_{\bm{\xi}}(\bm{\sigma}) = -c\left\langle
    \sum_{\bm{\sigma}^\prime,\bm{\sigma}^{\prime\prime}}P_{\bm{\xi}^\prime}(\bm{\sigma}^\prime)P_{\bm{\xi}^{\prime\prime}}(\bm{\sigma}^{\prime\prime})\left(
    e^{\frac{\beta}{c}(\bm{\xi}\bm{\xi}^\prime\bm{\xi}^{\prime\prime})(\bm{\sigma}\bm{\sigma}^\prime\bm{\sigma}^{\prime\prime})}-1
    \right)
    \right\rangle_{\bm{\xi}^\prime,\bm{\xi}^{\prime\prime}}+\Lambda_{\bm{\xi}},
    \label{eqn:P_variation}
\end{equation}
\end{widetext}
where $\Lambda_{\bm{\xi}}$ is a Lagrange multiplier ensuring the normalization constraint $\sum_{\bm{\sigma}}P_{\bm{\xi}}(\bm{\sigma})=1$. 

Combining this equation with the saddle-point condition obtained from variation with respect to $\hat{P}_{\bm{\xi}}(\bm{\sigma})$,      
we eventually obtain the self-consistent functional equation: 
\begin{equation}
    P_{\bm{\xi}}(\bm{\sigma}) = \frac{\exp\left(\Phi_{\bm{\xi}}(\bm{\sigma})\right)}{\sum_{\bm{\sigma}^\prime}\exp(\Phi_{\bm{\xi}}(\bm{\sigma}^\prime))},
    \label{eqn:spe_P}
\end{equation}
where 
\begin{widetext}
\begin{equation}
\Phi_{\bm{\xi}}(\bm{\sigma})
=
c
\left\langle
\sum_{\bm{\sigma}^\prime,\bm{\sigma}^{\prime\prime}}
P_{\bm{\xi}^\prime}(\bm{\sigma}^\prime)
P_{\bm{\xi}^{\prime\prime}}(\bm{\sigma}^{\prime\prime})
\left[
e^{\frac{\beta}{c}
(\bm{\xi}\bm{\xi}^\prime\bm{\xi}^{\prime\prime})
(\bm{\sigma}\bm{\sigma}^{\prime}\bm{\sigma}^{\prime\prime})}
-1
\right]
\right\rangle_{\bm{\xi}^\prime,\bm{\xi}^{\prime\prime}} .
\label{eqn:Phi}
\end{equation}
\end{widetext}
The free-energy density is evaluated by substituting the saddle-point solution, $P^*_{\bm{\xi}}(\bm{\sigma})$ into the action of Eq.~(\ref{eqn:free_energy_expression}).

The physical observables of the system can be obtained from the saddle-point order-parameter function $P^*_{\bm{\xi}}(\bm{\sigma})$. We first define the sublattice-resolved one-replica and two-replica moments as 
\begin{align}
m_{\bm{\xi}}^\alpha &= \frac{1}{|I_{\bm{\xi}}|}\sum_{i\in I_{\bm{\xi}}}\sigma_i^\alpha=\sum_{\bm{\sigma}} \sigma^\alpha P^*_{\bm{\xi}}(\bm{\sigma}), \\
q_{\bm{\xi}}^{\alpha\beta}
&=
\frac{1}{|I_{\bm{\xi}}|}
\sum_{i\in I_{\bm{\xi}}}
\sigma_i^\alpha\sigma_i^\beta = \sum_{\bm{\sigma}}
\sigma^\alpha\sigma^\beta P^*_{\bm{\xi}}(\bm{\sigma})
\qquad (\alpha\ne\beta).
\end{align}
Here, $m_{\bm{\xi}}^\alpha$ represents the average spin in replica $\alpha$ within the sublattice $I_{\bm{\xi}}$, while $q_{\bm{\xi}}^{\alpha\beta}$ is the replica overlap within the same sublattice. 
The corresponding global order parameters are obtained by averaging these sublattice-resolved quantities over the empirical sublattice fractions. Using the self-averaging property of $p_{\bm{\xi}}$ in Eq.~(\ref{eqn:self_averaging}), we obtain the pattern overlap $m^{\mu\alpha}$ and the spin-glass order parameter $q^{\alpha\beta}$ as 
\begin{align}
m^{\mu\alpha} &= \frac{1}{N}\sum_i\xi_i^\mu\sigma_i^\alpha=\left\langle \xi^\mu m_{\bm{\xi}}^\alpha\right\rangle_{\bm{\xi}},\\ 
q^{\alpha\beta} & = \frac{1}{N}\sum_i\sigma_i^\alpha\sigma_i^\beta=\left\langle
q_{\bm{\xi}}^{\alpha\beta}
\right\rangle_{\bm{\xi}}.
\end{align}
Thus, the global retrieval and spin-glass order parameters are determined by the sublattice-resolved moments encoded in the saddle-point order-parameter function $P^*_{\bm{\xi}}(\bm{\sigma})$. 

\subsection{Replica symmetric ansatz, local-field distribution and stability}
\label{sec:RS_local_field}
To proceed with the analysis of the saddle-point equations, we employ the replica symmetric (RS) ansatz. Under this ansatz, the order-parameter function $P^{\mathrm{RS}}_{\bm{\xi}}(\bm{\sigma})$ is invariant under any permutation of the replica indices. Since $\sigma^\alpha=\pm 1$, this implies that $P^\mathrm{RS}_{\bm{\xi}}(\bm{\sigma})$ depends on the replica spins only through their sum, 
\begin{align}
    P_{\bm\xi}^\mathrm{RS}(\bm\sigma) = F_{\bm\xi}\left(\sum_{\alpha=1}^n\sigma^\alpha\right). 
\end{align}
For finite-connectivity systems, we adopt the standard RS parametrization in terms of the local-field distribution $W_{\bm{\xi}}(h)$, expressed as  
\begin{equation}
P^\mathrm{RS}_{\bm{\xi}}(\bm{\sigma}) = \int dh W_{\bm{\xi}}(h) \frac{e^{\beta h \sum_{\alpha} \sigma^\alpha}}{(2 \cosh \beta h)^n},
\label{eqn:RS_ansatz}
\end{equation}
where $W_{\bm\xi}(h)$ represents the distribution of the effective field acting on spins in sublattice $I_{\bm\xi}$. 
Under this ansatz, the macroscopic order parameters become independent of the replica index. The sublattice magnetization $m_{\bm{\xi}}$ and the sublattice spin-glass order parameter $q_{\bm{\xi}}$ are given by the moments  
\begin{align}
    m_{\bm{\xi}} & = \int dh W_{\bm{\xi}}(h)\tanh(\beta h),\\
    q_{\bm{\xi}} & = \int dh W_{\bm{\xi}}(h)\tanh^2(\beta h). 
\end{align}
The corresponding global order parameters are the pattern overlap $m^\mu=\langle\xi^\mu m_{\bm{\xi}}\rangle_{\bm{\xi}}$ and the spin-glass order parameter $q=\langle q_{\bm{\xi}}\rangle_{\bm{\xi}}$. These quantities are used to distinguish the paramagnetic phase ($m^\mu=0, q=0$), the spin-glass phase ($m^\mu=0, q>0$), and the retrieval phase, in which at least one pattern overlap is nonzero.  

Substituting Eq.~(\ref{eqn:RS_ansatz}) into Eq.~(\ref{eqn:spe_P}) and taking the replica limit $n\to 0$, we obtain a self-consistent equation for $W_{\bm{\xi}}(h)$. Using the integral representation of the $\delta$-function and its Fourier transform, this equation can be expressed as  
\begin{widetext}
\begin{equation}
W_{\bm{\xi}}(h) = \int \frac{d\omega}{2\pi} e^{-ih\omega} \exp \left( c \left\langle \iint dh' dh'' W_{\bm{\xi}'}(h') W_{\bm{\xi}''}(h'') \left[ e^{i\omega u(\bm{\xi}, \bm{\xi}', \bm{\xi}''; h', h'')} - 1 \right] \right\rangle_{\bm{\xi}', \bm{\xi}''} \right),
\label{eqn:W_self_consistent}
\end{equation}
where $u$ is the effective field contribution generated by a single three-body interaction: 
\begin{equation}
u(\bm{\xi}, \bm{\xi}', \bm{\xi}''; h', h'') = \frac{1}{\beta} \tanh^{-1} \left[ \tanh\left( \frac{\beta}{c} \left(\bm{\xi} \bm{\xi}'\bm{\xi}''\right) \right) \tanh(\beta h') \tanh(\beta h'') \right].
\label{eqn:u_def}
\end{equation}
\end{widetext}
Equation~(\ref{eqn:W_self_consistent}) is the RS self-consistent equation for the local-field distribution of the finite-connectivity three-body Hopfield model.


The structure of the self-consistent equation (\ref{eqn:W_self_consistent}) has a simple statistical interpretation. Let $\chi_{\bm{\xi}}(k)=\int dh e^{ikh}W_{\bm{\xi}}(h)$ be the characteristic function of the local-field distribution. Equation (\ref{eqn:W_self_consistent}) is then equivalent to  
\begin{equation}
\chi_{\bm{\xi}}(k) = \exp \left( c \left[ \Psi_{\bm{\xi}}(k) - 1 \right] \right),
\label{eqn:characteristic_func}
\end{equation}
where 
\begin{equation}
\Psi_{\bm{\xi}}(k) = \left\langle \iint dh' dh'' W_{\bm{\xi}'}(h') W_{\bm{\xi}''}(h'') e^{ik u(\bm{\xi}, \bm{\xi}', \bm{\xi}''; h', h'')} \right\rangle_{\bm{\xi}', \bm{\xi}''}.
\end{equation}
The exponential form in Eq.~(\ref{eqn:characteristic_func}) is the characteristic function of a compound Poisson distribution. Therefore, the local field $h$ acting on a spin in sublattice $I_{\bm{\xi}}$ can be interpreted as a sum of Poisson-distributed numbers of independent three-body field contributions. This interpretation directly reflects the finite-connectivity structure of the model, in which each spin participates in $c$ connected triplets on average. 
The interpretation also provides a direct basis for the population dynamics algorithm used to solve the self-consistent equation numerically, as described in Sec.~\ref{subsec:Population_dynamics}. Alternatively, the same self-consistent equation can be derived within the replica-symmetric cavity framework by summing three-body messages independently on a locally tree-like random hypergraph\cite{Mézard2003, MezardMontanari}. 

This three-body structure also affects the stability analysis around the paramagnetic solution. In conventional two-body finite-connectivity models, transition points can often be identified by linearizing the self-consistent equations around the trivial solution. In the present model, however, the effective field $h$ is proportional to the product $\tanh(\beta h')\tanh(\beta h'')$. Consequently, the expansion of Eq.~(\ref{eqn:W_self_consistent}) around the paramagnetic solution $W_{\bm{\xi}}(h)=\delta(h)$ has no linear term in small field perturbation. Thus, within the RS local field equation, the paramagnetic solution does not lose stability through a linear instability. Nontrivial retrieval or glassy solutions must instead appear as finite-amplitude solutions, which makes a direct numerical solution of Eq.~(\ref{eqn:W_self_consistent}) essential for determining the phase diagram. 

\subsection{Replica symmetric free energy and stability of the RS solution}
\label{sec:RS_free_energy}
Nontrivial retrieval or glassy solutions appear as finite-amplitude fixed points of the RS local-field equation. Their thermodynamic relevance therefore cannot be determined from the local stability of the paramagnetic solution alone. Rather, it is necessary to compare the RS free-energy densities of competing fixed-point branches.  

Within the replica method under the RS ansatz, the RS free-energy density is obtained by evaluating the saddle-point action after eliminating the conjugate order-parameter functions and taking the replica limit $n\to 0$. In terms of the local-field distribution $W_{\bm{\xi}}(h)$ satisfying Eq.~(\ref{eqn:W_self_consistent}), the RS free-energy density is expressed as 
\begin{widetext}
\begin{align}
    f_{\mathrm{RS}} & = -\frac{1}{\beta} \left\langle \int dh W_{\bm{\xi}}(h) \ln \left[ 2 \cosh(\beta h) \right] \right\rangle_{\bm{\xi}}
    -\frac{c}{3\beta}\left\langle\ln\cosh\left(\frac{\beta}{c}(\bm{\xi}\bm{\xi}'\bm{\xi}'')\right)\right\rangle_{\bm{\xi}, \bm{\xi}', \bm{\xi}''}
    \nonumber \\
 &+  \frac{c}{\beta}\left\langle
 \int dh' dh'' W_{\bm{\xi}'}(h') W_{\bm{\xi}''}(h'') \ln \cosh\left(\beta u(\bm{\xi}, \bm{\xi}', \bm{\xi}''; h',h'')\right)\right\rangle _{\bm{\xi}, \bm{\xi}', \bm{\xi}''} \nonumber \\
 & + \frac{2c}{3\beta} \left\langle \iiint dh dh'dh''W_{\bm{\xi}}(h) W_{\bm{\xi}'}(h') W_{\bm{\xi}''}(h'') \right. \nonumber \\
& \quad \times \left. \ln \left[ 1 + \tanh\left( \frac{\beta}{c} (\bm{\xi}\bm{\xi}'\bm{\xi}'') \right) \tanh(\beta h) \tanh(\beta h') \tanh(\beta h'') \right] \right\rangle_{\bm{\xi}, \bm{\xi}', \bm{\xi}''}.
\label{eqn:free_energy_total}
\end{align}
\end{widetext}

The internal-energy density can also be evaluated directly from the same RS local-field distribution. 
Defining
\begin{equation}
    K(\bm{\xi},\bm{\xi}',\bm{\xi}'')=\tanh\left(\frac{\beta}{c}\left(\bm{\xi}\bm{\xi}'\bm{\xi}''\right)\right),
\end{equation}
the RS internal-energy density is given by 
\begin{align}
    e_{\mathrm{RS}} & = -\frac{1}{3}\left\langle
    \int dhdh'dh'' W_{\bm{\xi}}(h)W_{\bm{\xi}'}(h')W_{\bm{\xi}''}(h'')\right.\nonumber \\
    &\left.
    (\bm{\xi}\bm{\xi}'\bm{\xi}'')\frac{K+\tanh(\beta h)\tanh(\beta h')\tanh(\beta h'')}{1+K\tanh(\beta h)\tanh(\beta h')\tanh(\beta h'')}
    \right\rangle_{\bm{\xi},\bm{\xi}',\bm{\xi}''}.
    \label{eqn:RS_internal_energy}
\end{align}
The factor $1/3$ reflects the fact that the total number of connected triplets is $Nc/3$ on average, while the interaction strength of each connected triplet is normalized by $1/c$. 
The entropy density is then obtained from 
\begin{equation}
    s_{\mathrm{RS}} = \beta(e_{\mathrm{RS}}-f_{\mathrm{RS}}). 
\end{equation}

For the trivial paramagnetic solution, $W_{\bm{\xi}}(h)=\delta(h)$, the expressions reduce to 
\begin{equation}
    f^{\mathrm{PM}}_{\mathrm{RS}}=-\frac{1}{\beta}\ln 2-\frac{c}{3\beta}\left\langle\ln\cosh\left(\frac{\beta}{c}(\bm{\xi}\bm{\xi}'\bm{\xi}'')\right)\right\rangle_{\bm{\xi}, \bm{\xi}', \bm{\xi}''},
\end{equation} 
and  
\begin{align}
    e^{\mathrm{PM}}_{\mathrm{RS}} = -\frac{1}{3}\left\langle 
    \left(\bm{\xi}\bm{\xi}'\bm{\xi}''\right)
     K(\bm{\xi},\bm{\xi}',\bm{\xi}'')
    \right\rangle_{\bm{\xi},\bm{\xi}',\bm{\xi}''}.
    \label{eqn:PM_internal_energy}
\end{align}
By evaluating these quantities for the fixed-point solutions obtained from the RS local-field equation, candidate thermodynamic transition points within the RS approximation can be identified from crossings of the corresponding free-energy densities. 
These quantities are used to evaluate the thermodynamic consistency of the RS solution, and the internal energy provides a direct comparison with MCMC simulations. 

While the RS free-energy density provides a criterion for thermodynamic stability within the RS approximation, the RS solution must also be checked for stability against replica-symmetry-breaking (RSB) perturbations. In fully connected spin glasses, this corresponds to the stability of the RS saddle point against replicon fluctuations, leading to the de Almeida-Thouless (AT) criterion~\cite{deAlmeidaThouless1978}. For finite-connectivity disordered systems, an analogous perturbative test can be formulated as the linear stability of the RS fixed point in the space of local-field distributions~\cite{Montanari2003}. In the present RS formulation, we evaluate this stability by tracking the propagation of an infinitesimal local-field perturbation through the self-consistent equation, which leads to the AT-like eigenvalue given by  
\begin{align}
    \lambda_{\mathrm{AT}} & = 2c\left\langle\int dh'dh'' W_{\bm{\xi}'}(h')W_{\bm{\xi}''}(h'')\right.\nonumber \\
   &\left.  \left(\frac{K\tanh\beta h''(1-\tanh^2\beta h')}{1-(K\tanh\beta h'\tanh\beta h'')^2}\right)^2
    \right\rangle_{\bm{\xi}\bm{\xi}',\bm{\xi}''}. 
\end{align}
A value of $\lambda_{\mathrm{AT}}>1$ indicates a linear instability of the RS fixed point against spin-glass-type perturbations. In the present three-body model, however, nontrivial solutions can also appear through finite-amplitude instabilities associated with discontinuous transitions. Therefore, the condition $\lambda_{\mathrm{AT}}<1$ should be interpreted only as local stability within this perturbative criterion, not as definitive proof of the global validity of replica symmetry. Thus, the thermodynamic transition must be determined by comparing the free energies of the competing fixed-point solutions within the RS framework.

\section{Numerical analysis and phase structure}
\label{sec:results}
We next investigate the solutions of the replica-symmetric self-consistent equations using population dynamics (PD) and compare them with finite-size MCMC simulations with the population annealing (PA) method. We first describe the PD implementation. We then study two representative cases at fixed connectivity $c=10$: a low-load case, $P=2$, where the retrieval solution coexists with a glassy fixed point, and a high-load case, $P=8$, where the retrieval solution disappears, and the glassy instability becomes the relevant transition out of the paramagnetic phase. We finally summarize these results into the phase diagram in the $\alpha-T$ plane. 

\subsection{Population dynamics method}
\label{subsec:Population_dynamics}
Since the self-consistent equation of Eq~(\ref{eqn:W_self_consistent}) is a functional equation for the local-field distribution, it is generally difficult to solve analytically. We therefore solve it numerically by the population dynamics method~\cite{Mezard_Parisi_2001}. In our numerical implementation, we use the gauge symmetry associated with the target pattern and set $\xi_i^1=1$ for all nodes without loss of generality. After averaging over the non-condensed patterns, the sublattices are statistically equivalent with respect to the target pattern, and the local-field distribution can be represented by a single distribution $W_{\bm{\xi}}(h)$. This distribution is approximated by a population of $M$ fields, $\{h^1,\cdots,h^M\}$. We typically use $M=10^5$. 

The stochastic update of the population follows the compound Poisson structure of the RS local-field equation. The procedure is summarized as follows:   
\begin{enumerate}
    \item Initialization. The population of $M$ fields is initialized depending on the target branch to be explored: 
    \begin{itemize}
        \item \textit{Retrieval-biased initial condition}.
        All fields are set to a positive constant, typically $h^i=3.0$, to bias the population toward the retrieval solution. 
        \item \textit{Uninformative random initial condition}.
        Fields are drawn independently from a normal distribution with zero mean and small variance $\Delta$, $\mathcal{N}(0,\Delta^2)$, to probe the paramagnetic or glassy solution without introducing a macroscopic retrieval bias. 
    \end{itemize}  
    \item Iterative Updates.
    The following elementary update is repeated until the empirical field distribution converges to a fixed-point distribution.  
    \begin{enumerate}
        \item Draw a degree $k$ from a Poisson distribution with mean $c$. If $k=0$, this step is skipped. 
        \item For the $k$ three-body interactions occurring at the update site, $2k$ field values are to be drawn independently and uniformly from the population, and subsequently grouped into pairs of the form $(h_{a,1},h_{a,2})$ for $a=1,\ldots, k$. 
         \item To incorporate crosstalk from the remaining $P-1$ background patterns, we sample a site environment vector $\bm{\zeta}=(\zeta^2,\ldots,\zeta^P)\in\{-1,+1\}^{P-1}$. For each interaction $a$, we also sample independent pattern signs $\xi_{a,1}^\mu$ and $\xi_{a,2}^\mu$ for the two neighboring sites in the triplet, and compute $S_a=\sum_{\mu=2}^P \zeta^\mu\xi_{a,1}^\mu\xi_{a,2}^\mu$. The effective coupling is then $K_a=\tanh(\frac{\beta}{c}(1+S_a))$. 
        \item Compute the new field as  
        \begin{equation}
            h_{\mathrm{new}} = \frac{1}{\beta}\sum_{a=1}^k \tanh^{-1} \left[K_a\tanh(\beta h_{a,1})\tanh(\beta h_{a,2})\right]. 
        \end{equation}
        \item Replace a randomly chosen field in the population by $h_{\mathrm{new}}$. 
    \end{enumerate}
    \item Convergence Check. 
    At regular intervals, typically every $10^5$ elementary updates, we measure the magnetization $m=\langle \tanh(\beta h)\rangle$ and the spin-glass order parameter $q=\langle \tanh^2(\beta h)\rangle$. The iteration is terminated when the changes in both $m$ and $q$ from the previous check are smaller than the tolerance $\epsilon=10^{-5}$. 
    \item Measurement. 
    After convergence, physical quantities are evaluated using the final population. The AT-like eigenvalue $\lambda_\mathrm{AT}$ is also estimated by Monte Carlo integration over the converged population.  
\end{enumerate}
To account for the statistical uncertainty inherent in the population dynamics process, the standard errors of the observables are calculated by averaging over $N_\mathrm{run}=5$ independent runs. 

\subsection{Phase transitions at low load: $P=2$ and $c=10$}
\label{sec:phase_transition_p2}
To clarify the nature of the phases and phase transitions in the present model, we first examine the case with a small number of patterns. Specifically, we focus on $P=2$ at a fixed mean connectivity $c=10$, corresponding to a low load $\alpha=0.2$. 

\begin{figure}
    \centering
    \includegraphics[width=\linewidth]{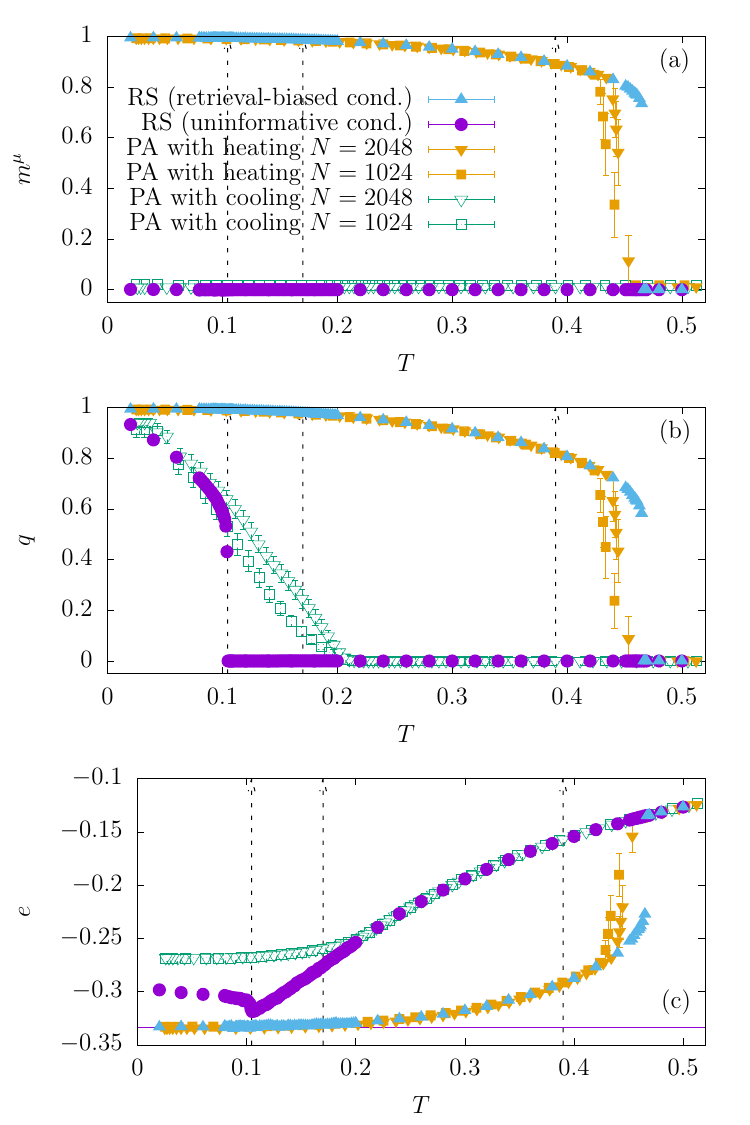}
    \caption{Temperature dependence of the macroscopic quantities for the three-body sparse Hopfield model with $c=10$ and $P=2$ ($\alpha=0.2$). Panels (a), (b), and (c) show the overlap magnetization $m^\mu$, the spin-glass (SG) order parameter $q$, and the internal energy density $e$, respectively. The replica-symmetric (RS) solutions obtained by population dynamics are shown as filled circles (uninformative random initial condition) and upward triangles (retrieval-biased initial condition). For comparison, microscopic simulation results obtained via population annealing (PA) are shown as open symbols for the cooling protocol and filled symbols for the heating protocol, using downward triangles ($N=2048$) and squares ($N=1024$). Error bars indicate sample-to-sample fluctuations in the PA results. The vertical dashed lines indicate, from left to right, $T_{\mathrm{SG}}^{\mathrm{RS}}$, $T_{s=0}$ and $T_c$, where $T_{\mathrm{SG}}^{\mathrm{RS}}$ is the onset temperature of the RS SG fixed point, $T_{s=0}$ is the temperature at which the RS entropy vanishes and below which it becomes negative, and $T_c$ is the first-order transition temperature estimated from the RS free-energy crossing. The horizontal line in panel (c) indicates the retrieval ground-state energy density, $-1/3$. }
    \label{fig:mq_c10p2}
\end{figure}

Figure~\ref{fig:mq_c10p2} shows the temperature dependence of $m^\mu$ and $q$, together with the internal energy density $e$. When the population dynamics (PD) for solving the self-consistent equation for $W(h)$ is initialized with an uninformative random condition, corresponding to the paramagnetic (PM) solution, the system remains in the PM state with $m^\mu=0$ and $q=0$ at high temperatures. As the temperature decreases, a finite spin-glass (SG) order parameter $q>0$ emerges from this random initialization at around $T\simeq 0.1$, while the pattern overlap $m^\mu$ remains zero, accompanied by a distinct change in the slope of the internal energy $e$. 
Conversely, when PD is initialized with a retrieval-biased condition, the retrieval state with $m^\mu>0$ and $q>0$ remains stable up to a higher temperature $T\simeq 0.45$, where it abruptly collapses into the PM fixed point. This upper limit of local stability defines the upper spinodal temperature $T_{\mathrm{spinodal}}$ of the retrieval solution. The coexistence of distinct RS fixed points over a finite temperature range is characteristic of a discontinuous transition. 

\begin{figure}
    \centering
    \includegraphics[width=\linewidth]{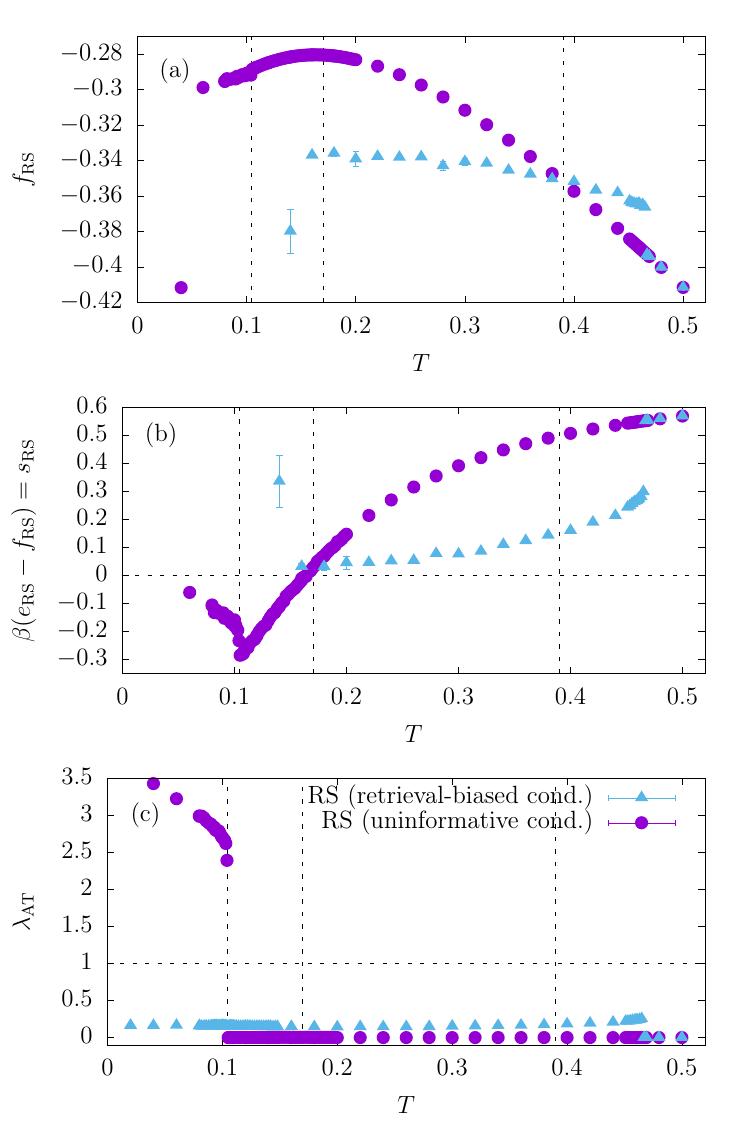}
    \caption{Temperature dependence of the thermodynamic quantities obtained from the replica-symmetric (RS) population dynamics for the three-body sparse Hopfield model with $c=10$ and $P=2$. The three panels display, from top to bottom, the free energy density $f_{\mathrm{RS}}$, the entropy density $s_{\mathrm{RS}}=\beta(e_{\mathrm{RS}}-f_{\mathrm{RS}})$, and the local stability measure $\lambda_\mathrm{AT}$.  The filled symbols represent the RS solutions obtained by population dynamics with uninformative random initial conditions (circles) and retrieval-biased initial conditions (triangles). The vertical dashed lines indicate, from left to right, $T_{\mathrm{SG}}^{\mathrm{RS}}$, $T_{s=0}$ and $T_c$, where $T_{\mathrm{SG}}^{\mathrm{RS}}$ is the onset temperature of the RS SG fixed point, $T_{s=0}$ is the temperature below which the RS entropy becomes negative, and $T_c$ is the thermodynamic first-order transition temperature determined by the crossing point of the RS free energies. }
    \label{fig:fl_c10p2}
\end{figure}

To investigate the physical validity and thermodynamic stability of these coexisting RS solutions, we examine the thermodynamic quantities derived from the fixed point of $W(h)$, as shown in Fig.~\ref{fig:fl_c10p2}. This figure plots the RS free-energy density $f_{\mathrm{RS}}$, the entropy density $s_{\mathrm{RS}}=\beta(e_{\mathrm{RS}}-f_{\mathrm{RS}})$, and the AT-like eigenvalue $\lambda_{\mathrm{AT}}$. The macroscopic hysteresis observed in the order parameters indicates a substantial discontinuous first-order phase transition. By comparing the free-energy densities of the two distinct branches in Fig.~\ref{fig:fl_c10p2}(a), the true thermodynamic transition is precisely located at $T\simeq 0.39$, indicated by the dashed vertical line, where the free-energy density of the ordered retrieval solution intersects that of the PM solution. 

As shown in Fig.~\ref{fig:fl_c10p2}(c), the AT-like eigenvalue of the retrieval branch remains consistently below unity, confirming that the retrieval phase is a robust local minimum of the free-energy landscape. Consequently, when the temperature increases beyond the RS free-energy crossing at $T_\mathrm{c}\simeq 0.39$, the retrieval state survives as a metastable branch up to the upper spinodal temperature $T_{\mathrm{spinodal}}\simeq 0.45$. Conversely, an analogous destabilization or a lower spinodal limit does not exist for the PM state when cooled from high temperatures. In this three-body interaction model, the pure PM state exhibits an identical local field $h_i=0$ at all sites, which causes the AT-like eigenvalue to be identically zero at any finite temperature. This property mathematically ensures that the PM fixed point remains stable against infinitesimal fluctuations down to $T=0$, preventing any continuous destabilization from the PM side within the RS ansatz.  

Despite the local stability of the PM fixed point against infinitesimal perturbations, another fixed point can be reached from a finite-amplitude random initial condition. Specifically, when the width $\Delta$ of the uninformative initial distribution is chosen large enough to cross the basin boundary of the PM fixed point, but remains sufficiently small to avoid introducing a macroscopic retrieval bias, the PD iteration escapes the PM fixed point and converges to an alternative SG-like solution with $q>0$ below $T\simeq 0.1$. Within a broad range of such initial widths, the resulting fixed-point distribution is insensitive to the precise value of $\Delta$. This behavior indicates that the SG-like solution has a sizable basin of attraction in the space of local-field distributions, although the PM solution remains locally stable in the linear sense. 

A closer examination of this SG-like RS fixed point reveals serious physical and thermodynamic pathologies. First, as shown in Fig.~\ref{fig:fl_c10p2}(c), the AT-like eigenvalue increases sharply above unity after the appearance of the $q>0$ solution, indicating a local instability of the RS fixed point against SG-type perturbations. Second, the internal-energy density of this branch, shown in Fig.~\ref{fig:mq_c10p2}(c), is not a monotonic function of temperature, which would imply an unphysical negative specific heat. Moreover, the entropy density in the PM branch becomes negative at $T\lesssim0.15$, before the finite-$q$ RS solution is reached at lower temperature. These thermodynamic inconsistencies indicate that the low-temperature PM branch is not physically described by the RS solution. Rather, they suggest that replica-symmetry breaking, possibly of one-step type, is required to describe the glassy state that emerges before the pathological RS SG fixed point is observed in the PD iteration. 

Finally, to support these theoretical insights, we compare them with MCMC simulations performed via population annealing (PA)~\cite{HukushimaIba, Machta2010} for the microscopic Hopfield model with finite system sizes. The detailed simulation setup is described in Appendix~\ref {sec:MCMC}. As shown in Fig.~\ref{fig:mq_c10p2}, the heating protocol of the PA simulations (filled symbols) exhibits excellent quantitative agreement with the RS retrieval solution for both the order parameters and the internal energy. It is noteworthy that, in the low-temperature limit, $T\to 0$, the energy density $e$ obtained from both the RS retrieval solution and the PA simulations converges precisely to $-1/3$. This value corresponds to the exact ground-state energy density of the pure retrieval phase, which is fundamentally independent of the pattern number $P$. The discrepancy near the upper spinodal temperature, where the order parameters jump abruptly, is consistent with finite-size effects and sample-to-sample fluctuations. The finite-size dependence is examined in more detail in Appendix~\ref{sec:MCMC}, where PA heating results show that the decrease in the retrieval magnetization systematically shifts toward the RS spinodal as the system size increases. Meanwhile, in the cooling protocol (open symbols), the PA simulation agrees with the trivial PM solution branch at higher temperatures and successfully reproduces the continuous emergence of the SG state in the low-temperature regime. Remarkably, while the unphysical RS-SG solution from the PD loses its thermodynamic monotonicity, the microscopic PA simulation yields a physically reasonable energy curve that decreases monotonically towards a low-energy state. 

The discrepancy between the pathological RS branch and the MCMC behavior suggests that the microscopic cooling process is not described by the RS-SG fixed point itself. Instead, it is more naturally interpreted as trapping in a rugged glassy landscape that requires an RSB description. Importantly, this glassy trapping should not be attributed solely to conventional crosstalk noise among the many embedded patterns. A closely related trapping phenomenon has been reported for a pure three-body ferromagnetic Ising model on regular random graphs~\cite{S.Franz2001}, which formally corresponds to the single-pattern case $P=1$ of the present model under a gauge transformation. Although the present model is defined on a Poisson random graph rather than a regular one, this comparison suggests that sparse three-body interactions can generate glassy metastable states even in the absence of conventional crosstalk. As a result, a cooling protocol starting from an uninformative initial condition can become trapped in glassy states rather than reaching the thermodynamically stable retrieval branch. 

\subsection{Phase transition at high load: $P=8$ and $c=10$}
\label{sec:phase_transition_p8}
\begin{figure}
    \centering
    \includegraphics[width=\linewidth]{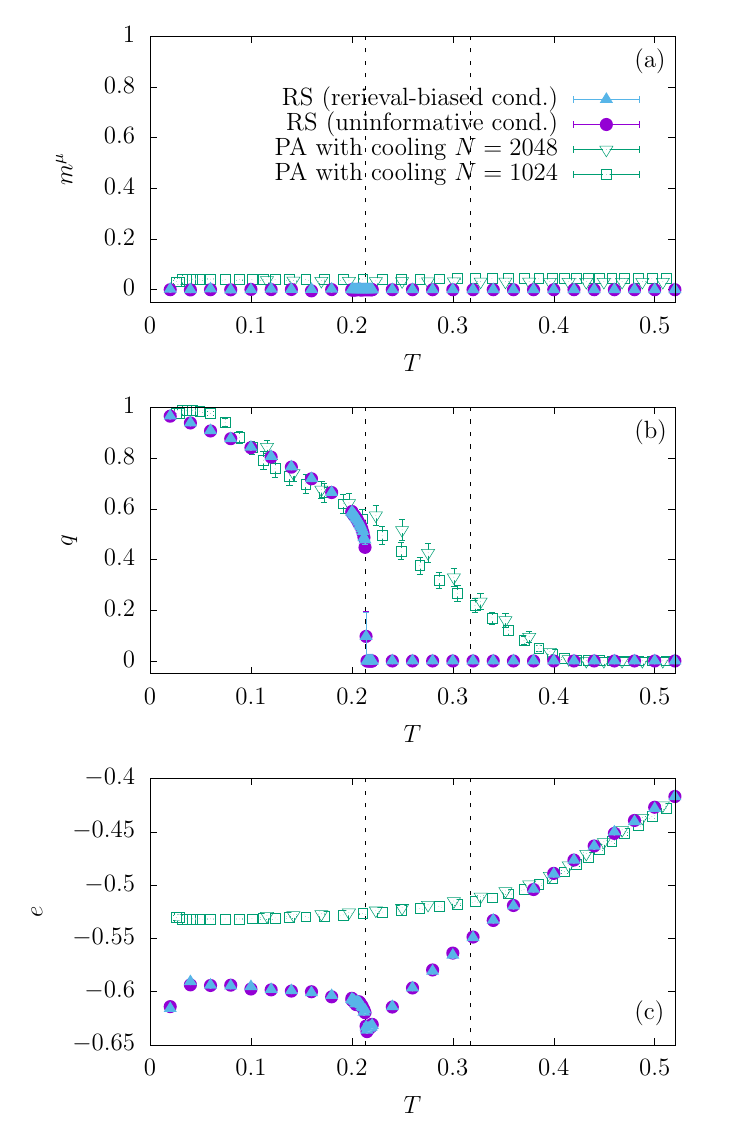}
    \caption{
    Temperature dependence of the macroscopic quantities for the three-body sparse Hopfield model with $c=10$ and $P=8$ ($\alpha=0.8$). The panels display, from top to bottom, the overlap magnetization $m^\mu$, the spin-glass (SG) order parameter $q$, and the internal energy density $e$. The filled symbols denote the replica-symmetric (RS) solutions, while the open symbols represent the microscopic PA simulation results for comparison. The vertical lines represent, from left to right, $T_{\mathrm{SG}}^{\mathrm{RS}}$ and $T_{s=0}$. 
    }
    \label{fig:mqe-c10p8}
\end{figure}

Next, we turn to the high-load regime by examining $P=8$ ($\alpha=0.8$) at the same mean connectivity $c=10$. In sharp contrast to the $P=2$ case, the retrieval solution with $m^\mu>0$ is no longer stable in this regime. Even when the population dynamics is initialized from a retrieval-biased condition, the pattern overlap rapidly collapses to $m^\mu=0$ at all temperatures investigated. We also examined the heating protocol in PA simulations, starting from the embedded memory patterns. In this high-load regime, however, these patterns have an energy density of $-1/3$ in the thermodynamic limit, significantly higher than that of the low-temperature glassy states. During the initial Monte Carlo updates, the system rapidly relaxes into lower-energy states, losing its overlap with the embedded patterns. Consequently, the heating protocol does not allow us to follow a metastable retrieval branch. As a result, no hysteresis is observed in the retrieval branch. We therefore present only the cooling results in Fig.~\ref{fig:mqe-c10p8}.

\begin{figure}
    \centering
    \includegraphics[width=\linewidth]{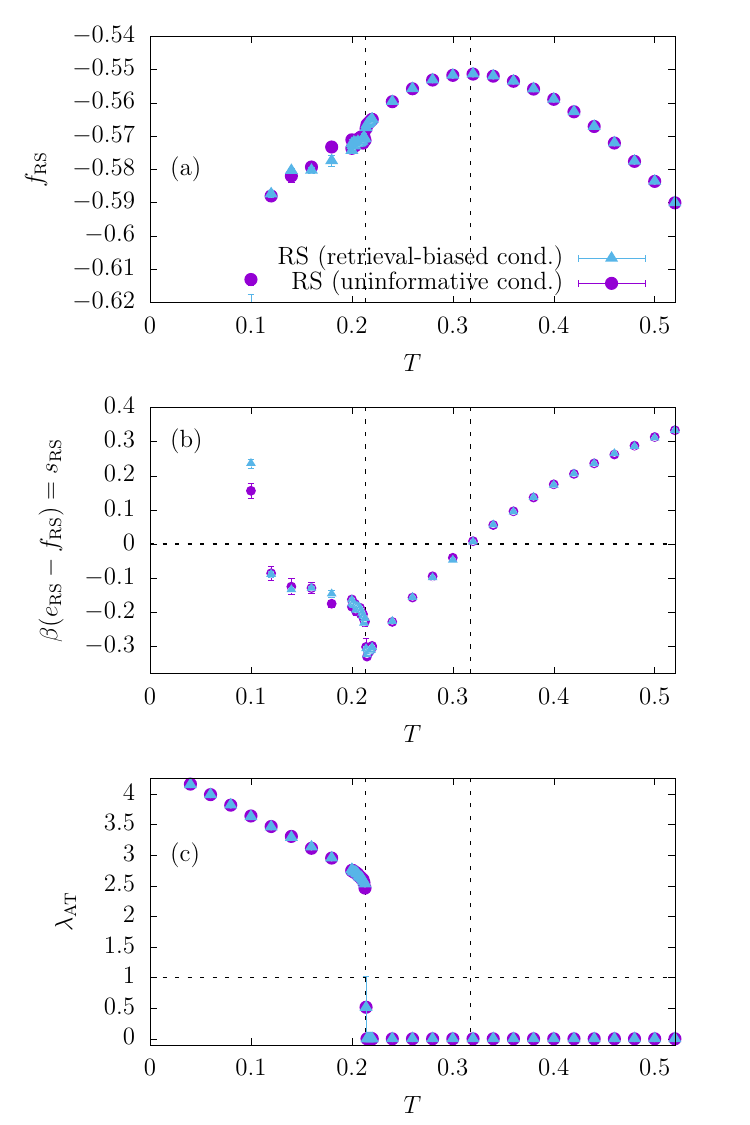}
    \caption{
    Temperature dependence of the thermodynamic quantities obtained from the replica-symmetric (RS) population dynamics for the three-body sparse Hopfield model with $c=10$ and $P=8$. The panels display, from top to bottom, the free-energy density $f_{\mathrm{RS}}$, the entropy density $s_{\mathrm{RS}}$, and the local stability measure $\lambda_{\mathrm{AT}}$. The vertical lines represent, from left to right, $T_{\mathrm{SG}}^{\mathrm{RS}}$ and $T_{s=0}$. 
    }
    \label{fig:fl-c10p8}
\end{figure}

Figures~\ref{fig:mqe-c10p8} and \ref{fig:fl-c10p8} show the temperature dependence of the macroscopic order parameters, the internal-energy density, and the corresponding thermodynamic quantities. At higher temperatures, the system remains in the trivial PM state with $m^\mu=0$ and $q=0$. In this regime, the internal-energy density already becomes lower than $-1/3$ at relatively high temperatures. Since $-1/3$ is the exact ground-state energy density of the pure retrieval state, this behavior indicates that at high load, the energy is dominated by competition among many embedded patterns rather than by a simple retrieval configuration. 

As the temperature is decreased, a nontrivial SG-like RS solution with $m^\mu=0$ and $q>0$ appears continuously at $T_{\mathrm{SG}}^{\mathrm{RS}}\simeq0.22$. However, this temperature should not be interpreted as the physical glass transition temperature. When the PM branch is followed from high temperature, the entropy density vanishes at $T_{s=0}\simeq 0.31$ and becomes negative below it, while the RS solution still remains in the formal $q=0$ PM state. Thus, the RS description must have broken down at a temperature higher than $T_{\mathrm{SG}}^{\mathrm{RS}}$, and the physical low-temperature phase should be replaced by an RSB glassy state before the pathological RS-SG branch appears. This interpretation is further supported by the AT-like stability analysis: at the point where the RS-SG branch appears, the eigenvalue $\lambda_{\mathrm{AT}}$ already exceeds unity, indicating that the branch is unstable against SG-type perturbations from its onset. 

As in the $P=2$ case, the direct PA simulations shown in Fig.~\ref{fig:mqe-c10p8} exhibit a clear deviation from the analytical RS predictions at low temperatures. Both the SG order parameter $q$ and the internal-energy density $e$ begin to deviate smoothly from the PM branch at temperatures higher than both the RS-SG onset $T_{\mathrm{SG}}^{\mathrm{RS}}\simeq 0.22$ and the negative-entropy onset $T_{s=0}\simeq 0.31$. This behavior supports the same physical picture inferred from the low-load case. Namely, the microscopic cooling dynamics is affected by a rugged glassy landscape at temperatures where the RS description is already thermodynamically inconsistent. The physically relevant low-temperature glassy state is therefore expected to require an RSB description rather than the pathological RS-SG fixed point.  

\subsection{Phase diagram at fixed connectivity $c=10$}
\label{sec:phase_diagram_c10}
\begin{figure}[t]
    \centering
    \includegraphics[width=\linewidth]{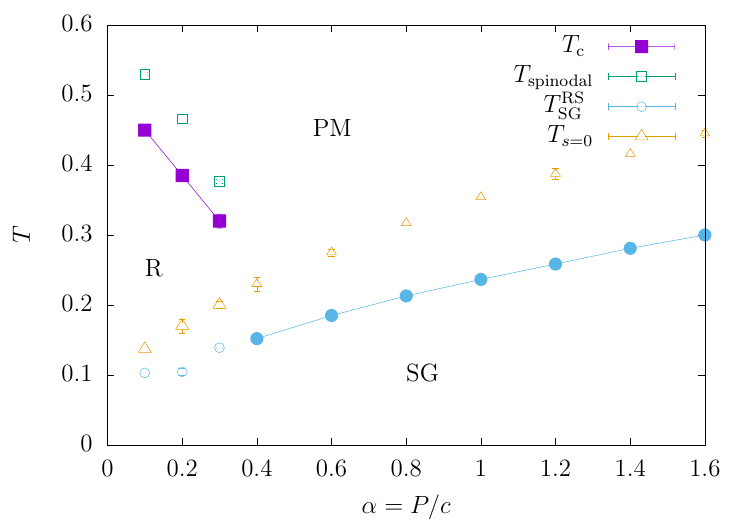}
    \caption{Replica-symmetric (RS) phase diagram of the three-body sparse Hopfield model for fixed mean connectivity $c=10$ in the $\alpha-T$ plane. The filled squares denote the first-order transition temperature $T_{\mathrm{c}}$ between the paramagnetic (PM) and retrieval (R) phases, determined by the crossing of the RS free energies. The open squares indicate the upper spinodal temperature $T_{\mathrm{spinodal}}$ of the retrieval state. The open circles represent the onset temperature $T_{\mathrm{SG}}^{\mathrm{RS}}$ of the SG-like fixed point in the low-load regime, where the PM branch is metastable with respect to the retrieval phase within the RS description. The filled circles indicate the corresponding onset temperature in the high-load regime, where the retrieval solution is absent. The open triangles indicate the temperature $T_{s=0}$ at which the entropy of the RS PM branch vanishes and below which it becomes negative, implying the breakdown of the RS description.  
   }
    \label{fig:phase_diagram_c10}
\end{figure}

To summarize the overall macroscopic behavior of the three-body sparse Hopfield model, we present the phase diagram on the $\alpha$-$T$ plane for fixed mean connectivity $c=10$ in Fig.~\ref{fig:phase_diagram_c10}. As demonstrated in the previous subsections, the model exhibits three characteristic macroscopic regimes. At sufficiently high temperatures, the system remains in the paramagnetic (PM) phase, regardless of the number of embedded patterns. In the low-load regime, a robust retrieval (R) phase appears at low temperatures, whereas in the high-load regime, the retrieval phase disappears and is replaced by a spin-glass (SG) phase. 

The phase diagram contains several characteristic temperatures associated with these phases, obtained from the replica-symmetric (RS) population dynamics (PD) analysis. The first-order transition temperature, denoted by $T_{\mathrm{c}}$ (filled squares), corresponds to the thermodynamic transition from the PM phase to the retrieval phase. The upper spinodal temperature of the retrieval state, denoted by $T_{\mathrm{spinodal}}$ (open squares), indicates the stability limit of the retrieval solution.  We also plot the temperature $T_{\mathrm{SG}}^{\mathrm{RS}}$ (circles) at which the SG solution appears in the PD iteration starting from a finite-amplitude perturbation of the PM solution, and the onset temperature $T_{s=0}$ at which the negative entropy emerges with decreasing temperature in the PM branch. 

The thermodynamic interpretation of these characteristic temperatures depends strongly on the load parameter $\alpha$. In the low-load regime, the RS free-energy analysis predicts a first-order transition from the PM to retrieval R phases at $T_{\mathrm{c}}$. Although the PM fixed point remains locally stable within the RS equations down to zero temperature, it becomes thermodynamically metastable below $T_{\mathrm{c}}$, and its formal continuation becomes physically inconsistent at $T_{s=0}$, where the RS entropy is negative. Therefore, the appearance of an SG-like fixed point at $T_{\mathrm{SG}}^{\mathrm{RS}}<T_{s=0}$ should not be interpreted as an equilibrium PM-to-SG phase transition. Rather, it reflects the structure of metastable solutions within the RS approximation. The PA simulations suggest that glassy behavior develops at higher temperatures than $T_{\mathrm{SG}}^{\mathrm{RS}}$, although its precise onset cannot be determined from the present finite-size cooling simulations. 

In contrast, in the high-load regime, the retrieval phase is absent, and the PM fixed point is thermodynamically stable at sufficiently high temperatures. However, as discussed above, the RS PM branch again develops negative entropy below $T_{s=0}$, before the SG-like fixed point appears at $T_{\mathrm{SG}}^\mathrm{RS}$. Thus, $T_{\mathrm{SG}}^{\mathrm{RS}}$ should not be identified with the physical PM-to-SG transition temperature in this regime either. The negative entropy indicates that the RS PM description must break down no later than $T_{s=0}$, while the SG-like RS solution itself exhibits an AT-like instability. These results suggest that an RSB treatment is necessary for a physically consistent description of the low-temperature glassy phase. The physical glass transition, if present, is expected to occur at or above $T_{s=0}$, but its precise location cannot be determined within the present RS analysis.    

Several remarks on the physical interpretation of this phase diagram are in order. First, because $P$ is an integer, the load parameter $\alpha=P/c$ takes discrete values for fixed finite connectivity $c$, even in the thermodynamic limit $N\to\infty$. Thus, for fixed $c=10$, a smooth critical storage capacity $\alpha_{\mathrm{c}}$ cannot be determined in the conventional sense as in fully connected models. This discreteness makes it difficult to determine the detailed structure of the phase diagram, particularly near the intersection or termination points of the retrieval and glassy branches. Nevertheless, the numerical results reveal a distinct feature of the three-body interaction model: a first-order transition to the retrieval phase in the low-load regime. In the PD analysis, this first-order nature appears as an initial-condition dependence of the fixed-point distribution, whereas in the population annealing (PA) simulations, it manifests as significant hysteresis between cooling and heating protocols. 

A particularly important feature is the subsequent state of the system when it is initialized in the high-temperature PM state and then cooled. Although the retrieval phase is thermodynamically stable in the low-load regime, the system can become trapped in a glassy state at low temperatures instead of reaching the retrieval state. This indicates a dynamical inaccessibility of the associative memory state from an uninformative initial condition. In other words, the system can retrieve a pattern when starting from one of the embedded patterns, but may fail to retrieve any pattern when cooled from an uninformative initial state. This behavior is qualitatively different from the continuous PM-to-retrieval transition found in the finite-connectivity two-body Hopfield model~\cite{wemmenhove2003Finite}. In the present three-body model, the PM fixed point remains locally stable within the RS description, while cooling from an uninformative state can lead to trapping in a complex glassy landscape generated by the three-body interactions. At high memory load, where the retrieval phase eventually disappears and the low-temperature RS PM branch becomes thermodynamically unphysical, these results strongly suggest a transition from the PM phase to a glassy phase. Determining the transition temperature and the thermodynamic properties of this phase requires an RSB analysis.  

\subsection{Mean-connectivity dependence of characteristic loads}
\label{subsec:connectivity_dependence}

\begin{figure}[t]
    \centering
    \includegraphics[width=\linewidth]{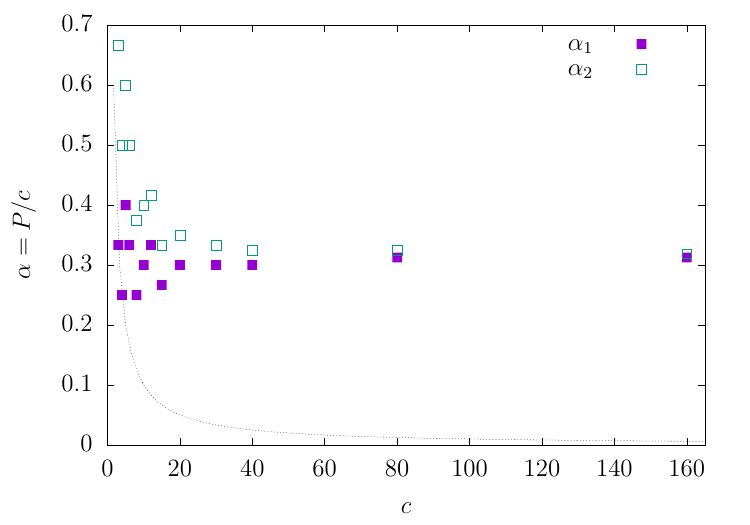}
    \caption{
    Mean-connectivity dependence of two characteristic load parameters obtained from the replica-symmetric self-consistent equations for the three-body sparse Hopfield model. The filled squares denote $\alpha_1$, the maximum load for which a stable retrieval fixed point is found at low temperatures. The open squares denote $\alpha_2$, the onset load of the high-load regime, where a spin-glass-like fixed point is obtained while no retrieval fixed point is found over the temperature range investigated. These two parameters are determined independently using distinct criteria and do not necessarily define a unique critical load. The dotted line indicates $\alpha=1/c$, corresponding to a single stored pattern $P=1$.  
    }
    \label{fig:c-alpha}
\end{figure}

Having examined the phase structure at fixed mean connectivity $c=10$, we next investigate how the characteristic load parameters obtained from the RS self-consistent equations depend on $c$. 
For this purpose, we define two load parameters based on distinct criteria: $\alpha_1$, the maximum load for which a retrieval fixed point remains stable at low temperatures, and $\alpha_2$, the onset load of the high-load regime, where an SG-like fixed point is found while no retrieval fixed point is observed over the temperature range investigated. These two quantities are determined independently and characterize different aspects of the change in the RS fixed-point structure as the load increases. Since $\alpha=P/c$ takes discrete values for fixed $c$, they do not necessarily define a unique critical load.  
Figure~\ref{fig:c-alpha} shows that both $\alpha_1$ and $\alpha_2$ exhibit the sawtooth-like dependence at small $c$, reflecting the discreteness of the number of embedded patterns $P$. The interval between $\alpha_1$ and $\alpha_2$ should not be interpreted as an additional PM phase, since these parameters are defined by different criteria for the RS fixed-point solutions rather than by thermodynamic phase boundaries. In the large-$c$ regime, however, both quantities approach finite values. This suggests that the characteristic number of storable patterns scales linearly with the mean connectivity, $P\propto c$, within the sparse regime. This scaling is natural from an information-theoretic viewpoint, since the number of nonzero couplings is proportional to $Nc$, but it is distinct from the fully connected three-body Hopfield model, where the storage capacity scales as $\mathcal{O}(N^2)$~\cite{AbbotArian1987}. Equivalently, if one considers a large but still sparse connectivity after taking the thermodynamic limit, the finite limiting values of $\alpha_1$ and $\alpha_2$ indicate that stable retrieval solutions persist up to loads of order $P=\mathcal{O}(c)$ within the RS description. 


\section{Summary and discussion}
\label{sec:summary}
In this study, we investigated the phase diagram of a three-body Hopfield model defined on sparse random graphs with finite mean connectivity $c=\mathcal{O}(1)$. By extending the functional replica framework of Wemmenhove and Coolen~\cite{wemmenhove2003Finite} to three-body interactions, we derived functional self-consistency equations for the local-field distribution, from whose solutions the replica-symmetric (RS) free-energy density can be evaluated. 
At the RS level, the same equations and thermodynamic quantities can also be obtained using the cavity method, which provides a more direct formulation in terms of local messages. 
The resulting equations were solved numerically by the population dynamics (PD) method and compared with finite-size MCMC simulations using population annealing. Our findings show that, in contrast to the continuous phase transitions observed in finite-connectivity two-body Hopfield models, three-body interactions give rise to discontinuous phase transitions, coexistence of the paramagnetic (PM) and retrieval solutions, and a clear spinodal structure. 

Another important feature of the present model is the behavior of the nonretrieval solutions at low temperatures. In the PD analysis, when the self-consistent equations are initialized from a finite-amplitude uninformative random distribution, the iterative update converges not to the retrieval solution but to an SG-like fixed point at low temperatures. A closely related glassy trapping phenomenon has been observed in Monte Carlo simulations of finite-connectivity three-body ferromagnets~\cite{S.Franz2001}. Indeed, for $P=1$, the present model reduces to a three-body ferromagnetic model under the gauge transformation $\tau_i=\xi_i\sigma_i$. Thus, glassy trapping can occur even in the absence of cross-talk noise among stored patterns, distinguishing it from the conventional cross-talk-induced glassiness of the standard two-body Hopfield model. Our PA simulations further show that similar trapping behavior persists for $P>1$, suggesting that the mechanism associated with sparse three-body interactions remains relevant even when multiple memory patterns are embedded.    

The scenario is qualitatively different from that of fully connected many-body Hopfield models. In the fully connected single-pattern limit, the model reduces to a mean-field three-body ferromagnet after a gauge transformation, which does not exhibit a glassy phase, although metastability associated with a first-order transition is possible. In contrast, finite-connectivity three-body models can exhibit glassy trapping even for a single stored pattern, as established by Franz et al.~\cite{S.Franz2001}. This distinction shows that glassy trapping does not necessarily require cross-talk among multiple stored patterns, but can instead arise from the combination of sparse connectivity and many-body interactions. Our results further show that this behavior persists when multiple memory patterns are embedded, while the RS analysis characterizes how the retrieval and SG-like fixed-point structures change with increasing memory load.  

We emphasize that the RS description of the SG-like branch should not be interpreted as a complete physical theory of the low-temperature glassy phase. As shown in Sec.~\ref{sec:results}, the RS analysis exhibits clear limitations at low temperatures: the formal PM branch develops negative entropy, while the SG-like fixed point exhibits a local instability signaled by an AT-like eigenvalue exceeding unity at sufficiently low temperatures. These features indicate the necessity of an RSB description of the glassy regime. Although 1RSB analyses of finite-connectivity many-body models and related random constraint-satisfaction problems have been developed in previous studies~\cite{Franz-PRL2001, Montanari_2008, MezardMontanari}, extending the present analysis beyond the RS level would require a more elaborate treatment of the local-field distributions in the presence of multiple stored patterns. Such an extension, which would be needed to determine the dynamical glass-transition temperature and characterize the low-temperature glassy phase, is beyond the scope of the present work. Nevertheless, the RS framework provides a useful reference theory for identifying candidate fixed-point branches, spinodal structures, and the qualitative structure of the phase diagram. In particular, the stability of the retrieval branch and its comparison with population annealing simulations are well captured by the RS population dynamics.  

In conclusion, our work demonstrates that finite-connectivity three-body Hopfield models exhibit discontinuous retrieval transitions, a distinct spinodal structure, and glassy trapping that persists in the presence of multiple stored patterns. As the load increases, retrieval fixed points eventually disappear, while SG-like solutions remain within the RS description. The characteristic number of storable patterns scales as $P=\mathcal{O}(c)$ with the mean connectivity. These results clarify the competition between retrieval and glassy behavior in higher-order associative memory models under sparse architectural constraints, leaving the detailed thermodynamic description of the low-temperature glassy phase for future work. 

\begin{acknowledgments}
This work was supported by JSPS KAKENHI Grant No.~23H01095 and JST Grant No.~JPMJPF2221.
The computations in this work were partially performed using the facilities of the Supercomputer Center, the Institute for Solid State Physics, the University of Tokyo (ISSPkyodo-SC-2024-Cb-0041, 2025-Ca-0081).
\end{acknowledgments}

\section*{data availability}
The numerical data underlying the results and figures presented in this paper are openly available in Zenodo at Ref.~\cite{dataset}.

\bibliography{sparseHopfield}

\appendix
\section{Details of population annealing Monte Carlo simulations}
\label{sec:MCMC}
\subsection{Population annealing simulation procedure}
\label{subsec:population_annealing}
To validate the theoretical results obtained from the replica analysis with microscopic simulations, we performed Markov-chain Monte Carlo (MCMC) simulations using the population annealing (PA) method~\cite{HukushimaIba, Machta2010}. PA is a population-based MCMC method suited to systems with rugged energy landscapes. In this method, a population of $R$ replicas is evolved under an adaptively determined temperature schedule, as described below, and each replica is assigned a statistical weight, which is uniformly distributed initially. When the inverse temperature is changed from $\beta$ to $\beta'$, the weight of each replica with energy $E$ is multiplied by the reweighting factor $\exp\left(-(\beta'-\beta)E\right)$, 
which is referred to as the Neal-Jarzynski factor~\cite{HukushimaIba}. The population is then resampled with replacement according to the normalized weights, after which the weights are reset to uniform values. Through this resampling step, replicas with larger weights tend to be duplicated, whereas those with smaller weights tend to be eliminated. After resampling, local updates by the Gibbs sampler are applied to each replica using a transition kernel that preserves the canonical distribution at the new temperature. 

In practice, the temperature schedule was chosen adaptively using the effective sample size (ESS) of the population. For a trial inverse-temperature increment $\Delta\beta$, the Neal-Jarzynski reweighting factors $w_r=\exp(-\Delta\beta E_r)$ were computed for each replica $r$, and the ESS was evaluated as 
\begin{align}
    \mathrm{ESS} = \frac{\left(\sum_{r=1}^R w_r\right)^2}{\sum_{r=1}^R w_r^2}. 
\end{align}
For the simulations presented in the main text, the next temperature was chosen so that $\mathrm{ESS}=0.8R$. We used a population size of $R=1024$ and performed $10^3$ Gibbs-sampling sweeps per replica at each temperature. To examine the sensitivity to the ESS target, we also performed simulations for $c=10$, $N=1024$, and $P=2$ and $8$ with $\mathrm{ESS}/R=0.65$ and $0.95$. The resulting energy densities showed little dependence on the ESS target over the temperature range investigated. This adaptive schedule automatically refines the temperature schedule in regions where the weights become broadly distributed, such as near phase transitions. 

We implemented two complementary protocols to probe different branches of the phase diagram.  
\begin{itemize}
    \item Cooling protocol: 
    The simulation is initialized at the high-temperature limit, $\beta=0$, where spin configurations can be sampled uniformly. The population is gradually cooled to the low-temperature regime.  
    \item Heating protocol: 
    In the low-load regime, the simulation is initialized from the embedded memory patterns and subsequently heated toward higher temperatures. Each replica is assigned one of $P$ embedded memory patterns with equal probability, and the population is then gradually heated toward higher temperatures.      
\end{itemize}
In principle, PA estimates thermodynamic expectation values by combining reweighting, resampling, and local MCMC updates. In the present three-body model, however, the strong first-order character of the transition leads to large free-energy barriers separating the distinct phases. As seen from the clear hysteresis in the numerical results, e.g., in Fig.~\ref{fig:mq_c10p2}, the population can remain trapped in long-lived metastable branches depending on the protocol. Therefore, the statistical averages obtained from our cooling and heating protocols should be interpreted as properties of these metastable branches rather than as guaranteed estimates of the global equilibrium states. 

To establish a direct comparison between the microscopic configurations obtained from PA simulations and the macroscopic quantities derived from the PD method, we define the finite-size order parameters as follows. In the analytical replica framework evaluated by PD, we use the ferromagnetic gauge to focus on a single retrieval pattern, taking $m=m^1$ and $m^\mu=0$ for $\mu\ge 2$ in the thermodynamic limit $N\to\infty$. In the PA simulations, by contrast, we do not impose any explicit gauge fixing to avoid artificially constraining the state space. 

For each sampled configuration $\bm{\sigma}^r=\{\sigma_i^r\}$ where $r=1,\cdots,R$, we first compute the $P$-dimensional overlap vector $\bm{m}^r=(m^{1,r},m^{2,r},\dots,m^{P,r})$,  with  
\[
m^{\mu,r}= \frac{1}{N}\sum_{i=1}^N\xi_i^\mu \sigma_i^r. 
\]
Since no gauge fixing is imposed in the PA simulation, the retrieval order parameter for replica $r$ is defined as the largest absolute pattern overlap,  
\begin{align}
    m^r & = \max_{\mu=1,\dots,P}|m^{\mu,r}|. 
\end{align}

To estimate the spin-glass order parameter corresponding to the RS quantity $q$, we use the overlap between two distinct replicas in the same disorder sample, 
\begin{align}
    q^{rs} & = \frac{1}{N}\sum_{i=1}^N\sigma_i^r\sigma_i^s. 
\end{align}
The spin-glass order parameter is then estimated from the root-mean-square replica overlap, 
\begin{align}
    q & = \left[\left\langle (q^{rs})^2\right\rangle_{r,s}\right]^{1/2},
\end{align}
where the average is taken over uniformly sampled replica pairs from the resampled population. 

After each resampling step, all replicas have equal statistical weight. Therefore, the expectation values of $m$ and $q$ can be estimated by averaging over the resampled population: $m$ is averaged over replicas, whereas $q$ is averaged over distinct replica pairs. The resulting estimates are averaged over independent PA runs and over 16-64 independent random realizations of the embedded patterns. Statistical errors are estimated as the standard errors of the disorder averages, calculated from the sample-to-sample fluctuations among these independent disorder realizations.   

Because of these definitions and the absence of gauge fixing, the measured order parameters contain finite-size effects. In the high-temperature paramagnetic phase, the spins fluctuate randomly, and each component of the pattern overlap vector scales as $m^\mu=\mathcal{O}(N^{-1/2})$. Consequently, the maximum value $m$ remains strictly positive at finite $N$. Similarly, the replica overlap $q^{rs}$ fluctuates on the scale $\mathcal{O}(N^{-1/2})$, so that the root-mean-square estimate of $q$ also exhibits an $\mathcal{O}(N^{-1/2})$ residual offset, even though both order parameters vanish in the thermodynamic limit. 

\subsection{Finite-size effect near the upper spinodal}
\label{subsec:finite_size_spinodal}
\begin{figure}
    \centering
    \includegraphics[width=\linewidth]{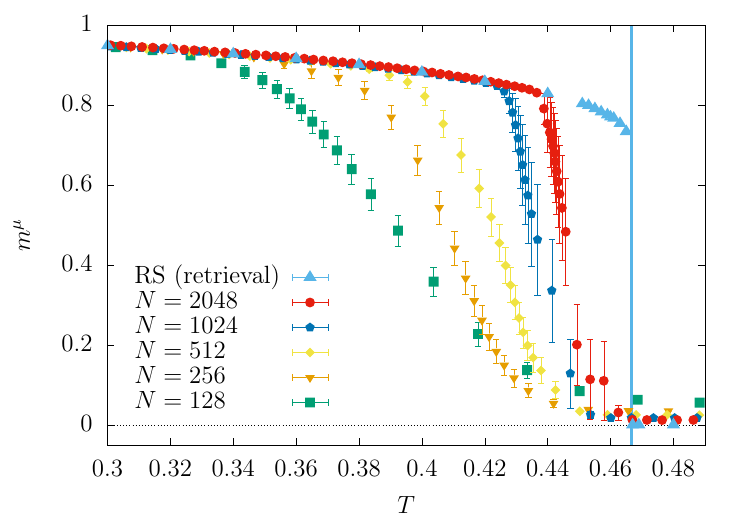}
    \caption{Finite-size dependence of the retrieval magnetization during population annealing with the heating protocol for $c=10$ and $P=2$. Results for $N=128$, $256$, $512$, $1024$, and $2048$ are compared with the RS solutions (upper triangles). The vertical line indicates the upper spinodal temperature of the RS retrieval solution. }
    \label{fig:finite_size_spinodal}
\end{figure}

To examine the finite-size dependence of the PA heating results near the upper spinodal of the retrieval solution, we compare the retrieval magnetization for different system sizes in Fig.~\ref{fig:finite_size_spinodal}. As the system size increases, the decrease in the retrieval magnetization becomes sharper and shifts systematically toward the upper spinodal temperature predicted by the RS analysis. Although no quantitative extrapolation to the thermodynamic limit is performed, this trend is consistent with the finite-size origin of the discrepancy between the PA results and the RS prediction near the spinodal. 

\end{document}